\documentclass[draft,showpacs,preprintnumbers,amsmath,amssymb]{revtex4-1}
\usepackage[utf8]{inputenc} 
\usepackage{array}
\begin{document}
\title{Superradiance instability of small rotating AdS black holes in arbitrary dimensions}
\author{\"Ozg\"ur Delice}
\email{ozgur.delice@marmara.edu.tr}
\affiliation{Physics Department, Marmara University, Faculty of Science and 
Letters, 34722 Istanbul, Turkey}

\author{T\"urk\"uler Dur\u gut}
\affiliation{Physics Department, Marmara University, Faculty of Science and 
Letters, 34722 Istanbul, Turkey}
\date{\today}
\begin{abstract}

We investigate the stability of $D$ dimensional singly rotating Myers-Perry-AdS black holes under superradiance against scalar  field perturbations. 
It is well known that small four dimensional rotating or charged Anti-de Sitter (AdS) black holes are unstable against superradiance instability of a scalar field. Recent works extended the existence of this instability to five dimensional rotating charged  AdS black holes or static charged AdS black holes in arbitrary dimensions. In this work we analytically prove that, rotating small AdS black holes in  arbitrary dimensions also show superradiance instability irrespective of the value of the (positive) angular momentum quantum number. To do this we solve the Klein-Gordon equation in the slow rotation, low frequency limit. By using the asymptotic matching technique, we are able to calculate the real and imaginary parts of the correction terms to the frequency of the scalar field due to the presence of the black hole, confirming the presence of superradiance instability.  We see that, unlike in the case of static AdS black holes, the analytical method is valid  for rotating AdS black holes for any value of angular momentum number and space-time dimensions. For comparison we derive the corresponding correction terms for Myers-Perry black holes  in the black hole bomb formalism  in Appendix and see that the results are in  agreement. 
\end{abstract}
\maketitle
\section{Introduction}

Superradiance is a phenomenon  where a field satisfying certain conditions is amplified by its interaction with a dissipative system \cite{review}. In the black hole superradiance \cite{Zeldowich,Bekenstein}, the interaction of scalar, electromagnetic or gravitational fields with the black hole event horizon, which behaves as a  one way membrane mimicking a dissipative system,  causes the field to be amplified, if the field satisfies the superradiance condition. 
The condition for superradiance for Kerr black holes  is  $\omega < m \Omega_h$  \cite{Bardeen,Starobinsky,Starobinsky1} where $\omega$ is the frequency of the wave, $m$ is the azimuthal number, and $\Omega_h$ is the angular velocity of the horizon of the black hole. A possible indirect observation of superradiant scattering from black hole--pulsar  binary systems were proposed  recently \cite{Rosa1}.
 The physics of superradiance and several aspects of  black hole superradiance  is discussed in an excellent review \cite{review} and we refer this review for further information.

The superradiance instability  requires  a mechanism such as a potential barrier, which localizes the field near the horizon of the black hole and does not permit scattered waves to escape to infinity. The continuous reflection of the amplified waves from the barrier   towards horizon  may lead to  an instability of the black hole, i.e., its rotational energy and angular momentum is decreased by such a process. 
It is well known that the Schwarzschild black holes are stable \cite{ReggeWheeler,Vishveswara} against such perturbations.  More general black holes, however,  may develop instabilities under certain conditions. For example, although the Kerr black hole is stable under massless scalar, electromagnetic or gravitational perturbations \cite{PressTeukolsky,PressTeukolsky1}, a massive scalar field may cause superradiant instability on this black hole \cite{Damour,Zouros,Detweiler,Furuhashi,Dolan1}, since the mass term  acts as a natural mirror.
In order to imitate the conditions leading to  the superradiance instability, the black hole bomb mechanism is often used \cite{zeldovich1,PressTeukolskyN,CardosoBomb,Strafuss,Dolan, HodHod,Rosa,Witek,AD1,Lee,Turkuler,Herdeiro1,Degollado,Hod,Li,Li1,Aliev5d}. In that mechanism the black hole is though to be surrounded by a  hypothetical  reflective mirror leading the field to be amplified back by the black hole, resulting an instability. The use of this hypothetical setting helps  to analyze certain aspects of this instability such as the time scale of the instability and the relation of the instability with the parameters of the black hole, the  field or the mirror. 

Although their astrophysical importance is limited, the black holes living in an AdS spacetime have become very important after the gauge/gravity duality is introduced \cite{Maldacena1}. Thus, it  has become a necessity to explore  the stability of these black holes under different perturbations, such as the superradiance instability.  
Since the boundary of AdS spacetime behaves as a mirror wall, a black hole living in an AdS spacetime may be unstable against superradiance as well, similar to the black hole bomb mechanism.
 Actually, it is well known that large AdS black holes are stable \cite{HawkingReal}. However, the four dimensional rotating  small \cite{CardosoDias,Uchikata} or charged \cite{Gubser1} AdS black holes are unstable against superradiance instability of a scalar or charged scalar field. In five dimensions, it was shown that small charged AdS black holes with two rotation parameters are also  unstable \cite{AD}. This instability is also reported for five dimensional hairy AdS black holes in \cite{Dias1}. Superradiance  also leads to gravitational instabilities for AdS black holes in four and higher dimensions \cite{Yoshida,KunduriLucietti,Kodama,Murata,Kodama1,Lehner}.
 Recently, small charged static AdS black holes in $D$ dimensions are also shown to be unstable \cite{WangHerdeiro} against superradiance, together with the correction of an erroneous conclusion of \cite{AD} that  for certain values of orbital number $l$, the instability is not triggered in five dimensions. Actually, this  is corrected   using a numerical approach,  since for those particular values of $l$,  analytical methods fail  for static black holes as in their case. See also \cite{Aliev5d} for five dimensional analytical treatment of such values of $l$ in the black hole bomb mechanism. In this paper, our aim is to investigate the stability of \emph{rotating} AdS black holes in arbitrary dimensions against scalar perturbations using the analytical methods. We will especially show that, for rotating black holes, unlike static ones, the analytical method is applicable for any (positive) value of orbital quantum number.  We also compare these results with the superradiance instability of $D$ dimensional singly rotating Myers-Perry black holes under scalar perturbations in the black hole bomb mechanism   which we review in the Appendix and find a perfect agreement. 

\section{Superradiance instability of Myers-Perry-AdS Black Holes}

\subsection{Metric and its properties}
A  general $D=4+n$ dimensional rotating-AdS black hole is given by the Myers-Perry-AdS solution \cite{Carter,Gibbons1,Gibbons2}. Here we consider the special case where only a single  nonvanishing rotation parameter $a$  \cite{HawkingHunter,Carter}  exists. This spacetime is described by
\begin{equation}
 ds^2=-\frac{\Delta_r}{\Sigma}\left(dt- a \frac{\sin^2\theta}{\Xi} d\phi \right)^2+\frac{\Sigma}{\Delta_r}dr^2+\frac{\Sigma}{\Delta_\theta}d\theta^2 + \frac{\Delta_\theta\sin^2\theta}{\Sigma}\left(a dt-\frac{r^2+a^2}{\Xi}d\phi \right)^2+r^2\cos^2\theta d\Omega_n^2,
\end{equation}
 where $M$ is the mass parameter,  $\ell=\sqrt{-(D-2)(D-1)/(2\Lambda)}$ is the AdS radius, $\Lambda$ is  a negative cosmological constant, and $d\Omega_n^2$ is the standard metric of the $n$ dimensional unit sphere. The metric functions are given by
\begin{eqnarray}
 &&\Sigma=r^2+a^2\cos^2\theta, \quad \Delta_r=(r^2+a^2)\left(1+\frac{r^2}{\ell^2} \right) -2 M r^{1-n},\\
&& \quad \Delta_\theta=1-\frac{a^2}{\ell^2}\cos^2\theta, \quad
\Xi=1-\frac{a^2}{\ell^2}.
\end{eqnarray}

This black hole has physical mass $\mu$ and angular momentum $J$ given by \cite{Gibbonsetal,Caldarelli}
\begin{eqnarray}\label{MJ}
\mu=\frac{A_{D-2}}{4\pi  \Xi^2}\left(1+\frac{D-4}{2}\,\Xi \right)M,\quad J=\frac{A_{D-2}}{4\pi  \Xi^2}M a,\quad A_{D-2}=\frac{2\pi^{(D-1)/2}}{\Gamma\left(\frac{D-1}{2}\right)},
\end{eqnarray}
where $A_{D-2}$ is the surface area of a unit $D-2$ sphere. The angular velocity of the horizon with respect to rotating infinity is given by
\begin{eqnarray}\label{angvel}
\Omega_h=\frac{\Xi a}{r_h^2+a^2},      
\end{eqnarray}
and corresponding expression  for the nonrotating infinity relevant for black hole   thermodynamics  is $\Omega=\Omega_h+a/\ell^2$.
The event horizon  of this black hole, $r=r_h$, is located at the largest root of the equation $\Delta_r(r)=0$. There is a Bogomol'nyi--Prasad--Sommerfield (BPS)--like upper bound \cite{BPSbound} on the rotation parameter, i.e. $|a|<\ell$, otherwise the metric describes a naked singularity. These black holes also suffer from  an ultraspinning instability \cite{Dias2,Gwak}.  It was demonstrated in \cite{Yoshida} for four dimensional Kerr-AdS black holes that superradiance instability  is present if the horizon angular velocity satisfies $\Omega \ell<1$ and the end point of the instability  is described by a Kerr-AdS black hole whose
boundary is an Einstein universe rotating with the speed of light or equivalently the corotating Killing vector becomes space-like and in equilibrium with  a scalar field cloud.

\subsection{Klein-Gordon equation}

In this subsection we consider the Klein-Gordon equation for a scalar field $\Phi$  with mass  $\tilde m$ of the form
\begin{equation}
\nabla^\mu \nabla_{\mu} \Phi-\tilde m^2 \Phi=0.
\end{equation}    
Since this equation is known to be separable \cite{Carter,Kunduri,Vasudevan} for general Myers-Perry-AdS black holes, we can consider the following  ansatz for the scalar field 
\begin{equation}
\Phi=e^{i m \phi-i\omega t} Y(\Omega)\Theta(\theta)R(r).
\end{equation}
The resulting equation separates into its angular and radial parts as follows:
\begin{eqnarray}\label{angular}
&&\frac{1}{\cos^n\theta \sin\theta}\frac{d}{d\theta}\left(\Delta_{\theta}\cos^n\theta 
\sin\theta \frac{d\Theta}{d\theta}\right) \nonumber
\\
&&\quad \quad -\bigg[\frac{1}{\Delta_\theta}\left(a\omega\sin\theta-\frac{\Xi m}{\sin\theta}\right)^2 + \frac{j(j+n-1)}{\cos^2 \theta}-\lambda_{jlm}-\tilde m^2a^2\cos^2\theta \bigg]\Theta =0,
\end{eqnarray}
\begin{eqnarray}\label{radial}
\frac{1}{r^{n}}\frac{d}{dr}\left(\Delta_r r^n \frac{dR}{dr}\right)
+\bigg[\frac{(a^2+r^2)^2}{\Delta_r}\left(\omega-\frac{m\Xi  a}{a^2+r^2}\right)^2-\frac{
j(j+n-1)a^2}{r^2}-\lambda_{jlm}-\tilde{m}^2 r^2\bigg]R=0.\ \ \ \
\end{eqnarray}

Here the terms $j(j+n-1)$ are the eigenvalues of the spheroidal equation for $Y$ of the $n$-sphere \cite{Muller} with $j$ assuming integer values. We need the eigenvalues $\lambda_{jlm}$ of the angular equation (\ref{angular}) where the eigenvalues of this equation for four \cite{Suzuki}, five \cite{AD} and higher dimensions \cite{Cho} is discussed  recently. In the slow rotation low frequency limit, i. e. $a \omega \ll 1 $, and for the case the mass of the scalar field is very small such that $\tilde m^2a^2\sim 0$, the eigenvalues $\lambda_{jlm}$ of  the  Kerr-AdS spheroidal harmonics    can be  expanded into a Taylor series   
as
\begin{equation}\label{lambda}
\lambda_{jlm}=l(l+n+1)+ \sum_{p=1}^{\infty}\tilde{f}_p \left(a \bar{\omega} \right)^p, 
\end{equation}
where $\bar{\omega}=\sqrt{\omega^2-\tilde m^2}$, $l$ is a positive integer which satisfies $l\ge j+|m|$, and certain terms of $\tilde{f}_p$  are given explicitly in \cite{Suzuki,Cho}.
 We take $m$ as a positive integer in this paper. The crucial  point here is that  $\lambda_{jlm}$ have small correction terms added to $l$ terms, which prevent the eigenvalues from being an exact integer, unless the rotation parameter $a$ vanishes, as in the case of static black holes.
  The radial equation cannot be solved analytically but using the matched asymptotic expansion method, presented first by Starobinsky \cite{Starobinsky}, we can obtain the asymptotic solutions near the black hole horizon and  in the far region. Their matching  at the intermediate region will enable us to calculate the correction term to the frequency of the wave and subsequently determine the existence of the instability.

\subsection{Asymptotic behaviour of the scalar field and the superradiance condition} 
In order to discuss the asymptotic behavior of the scalar field, it is useful to introduce a tortoise coordinate $r^*$  by considering the transformation $r=r(r^*)$ of the radial coordinate and the wave function as follows:
\begin{eqnarray}\label{tortR}
\mathcal{R}(r)&=&\sqrt{r^n(r^2+a^2)}R,\\
\frac{dr^*}{dr}&=&\frac{r^2+a^2}{\Delta_r}. \label{tortr}
\end{eqnarray}
These transformations bring the radial equation into a Schr\" odinger--like form,
 \begin{equation}
 \frac{d^2\mathcal{R}}{dr^{*2}}+V(r)\mathcal{R}=0,
 \end{equation}
 where the potential term $V(r)$ is given by
 \begin{eqnarray}
 V(r)=&&\left(\omega-\frac{\Xi m a}{r^2+a^2} 
\right)^2+\frac{\Delta_r}{(r^2+a^2)^2}\bigg\{\frac{j(j+n-1)a^2}{r^2} +\lambda_{jlm}+\tilde{m}^2r^2  \nonumber \\ &&
\quad \quad \quad  \quad \quad \quad \quad \quad \quad \left.  +
\frac{\sqrt{r^2+a^2}}{r^{n/2}}\frac{d}{dr}\left[ r^n \Delta_r 
\frac{d}{dr}\frac{1}{\sqrt{r^n(r^2+a^2)}} \right] \right\}. \label{tortpot}
\end{eqnarray} 

Near the horizon $r\rightarrow r_h$, since $\Delta_r\sim 0$, the potential term 
becomes 
\begin{equation}
V(r\rightarrow r_h)\sim (\omega-m \Omega_h)^2. \label{potnearhor}
\end{equation}
 This implies that  the solution near horizon becomes, in the classical limit where only the ingoing waves present,  $R(r\rightarrow r_h)\sim e^{-i \omega t-i(\omega-m \Omega_h) r^*}$ and that  the superradiance condition is
\begin{equation}
\omega<m \Omega_h,
\end{equation}
where the field is amplified when the frequency of the wave satisfies this condition.

As $r\rightarrow \infty$ the potential becomes infinitely large, implying the vanishing of the scalar field at radial infinity, i.e.
\begin{equation}
\Phi(r\rightarrow \infty)\rightarrow 0.
\end{equation}
As a consequence of these observations, the appropriate boundary conditions turn out to be  the Dirichlet boundary condition at the radial infinity and incoming wave boundary conditions on the horizon of the black hole.

 \subsection{Near region solution}
 
Here we solve the radial part of the Klein-Gordon equation at the region near the event horizon of the black hole $r\sim r_h$ in the slow rotation low frequency limit, noting that near the horizon the effects of the cosmological constant are negligible.  We also assume that the Compton wavelength of the perturbations are much larger than the size of the horizon. i. e., $\tilde m r_h\ll 1$ . In 
these limits, i. e., $r-r_h\ll\frac{1}{\omega}$, $\omega a \sim 0$,  $a^2\sim 0$, $r_h\ll \ell$, $a\ll \ell$,   
Eq. 
(\ref{radial})  becomes
\begin{eqnarray}\label{radnearads}
r^{-n}\frac{d}{dr}\left(\Delta_r r^n \frac{dR}{dr}\right)
+\bigg[\frac{r_h^4}{\Delta_r}\left(\omega-m\Omega_h\right)^2- \zeta\bigg]R(r)=0,
\end{eqnarray}
where
\begin{equation}\label{lambda1}
\zeta=\lambda_{jlm}+\tilde m^2r_h^2
\approx l(l+n+1)+\epsilon.
\end{equation}
Note that the term $\epsilon$ carries all the correction terms due to the rotation of the black hole and  mass of the scalar field.  This small term is not arbitrary and can be calculated using  (\ref{lambda}) for given values of the parameters of the black hole and the scalar field.  We will keep this term since it will be crucial in the forthcoming analysis. 
In order to obtain solutions of Eq. (\ref{radnearads}), let us consider a new radial 
coordinate
\begin{eqnarray}
x=r^{n+1},
\end{eqnarray}
which brings the radial equation into
\begin{eqnarray}\label{xradial}
(n+1)^2\bar{\Delta}\frac{d}{dx}\left(\bar{\Delta}\frac{dR}{dx}\right)
+\bigg[x^{\frac{2(n+2)}{n+1}}_+\left(\omega-m\Omega_h\right)^2-\zeta\,\bar{
\Delta}\bigg]R(r)=0,
\end{eqnarray}
where $\bar{\Delta}=r^{2n}\Delta_r \approx x^2-2Mx+a^2x^{2n/(1+n)}=(x-x_+)(x-x_-)$. In four dimensions $n=0$,  we have two horizons, and in the higher dimensions we can set the last term in the approximation and the inner horizon to zero. Nevertheless, in order to have a unified treatment we keep the $x_-$ term in our expressions for higher dimensions as well. 
To bring Eq. (\ref{xradial}) into the form of a hypergeometric 
equation, we first  define a new dimensionless variable
\begin{eqnarray}
z=\frac{x-x_+}{x-x_-},
\end{eqnarray}
with the event horizon being at $z=0$. Then the radial equation becomes
\begin{eqnarray}\label{zradial}
z(1-z)\frac{d^2R}{dz^2}
+(1-z)\frac{dR}{dz}
+\left[\frac{1-z}{z}\tilde \omega^2-\frac{1}{1-z}\tilde \zeta\right]R=0,
\end{eqnarray}
where 
\begin{eqnarray}
&&\tilde \omega =\frac{x^{\frac{n+2}{n+1}}_{+}(\omega-m\Omega_h)}{(n+1)(x_+-x_-)}, \label{omegatilde} \\
&& \tilde\zeta=\frac{\zeta}{(n+1)^2}.
\end{eqnarray}
Finally, if we define the radial part of the wave function $R$ in terms of a new 
function $\mathcal{F}(z)$ as follows
\begin{eqnarray}\label{RFtrans}
R=z^{i\tilde \omega}(1-z)^{\frac{1+\sqrt{1+4\tilde \zeta}}{2}}\mathcal{F}(z),
\end{eqnarray}
we see that the radial part of the Klein-Gordon equation (\ref{zradial})  is 
equivalent to a hypergeometric differential equation given below:
\begin{eqnarray}\label{hypergeometric}
z(1-z)\frac{d^2 \mathcal{F}}{dz^2}
+\left[\gamma-\left(\alpha+\beta+1\right)z\right]\frac{d \mathcal{F}}{dz}
-\alpha\beta \mathcal{F}=0,
\end{eqnarray}
where the constant parameters $\alpha,\beta$ and $\gamma$ are given by
\begin{eqnarray}
\alpha&=&\frac{1+\sqrt{1+4\tilde \zeta}}{2}+2\,i\,\tilde \omega ,\nonumber\\
\beta&=&\frac{1+\sqrt{1+4\tilde \zeta}}{2},\\
\gamma&=&1+2i\, \tilde \omega. \nonumber
\end{eqnarray}
The most general solution of this equation in the neighborhood of $z=0$ is  given by 
\begin{eqnarray}
\mathcal{F}(z)=A\,z^{1-\gamma}F(\alpha-\gamma+1,\beta-\gamma+1,2-\gamma,z)+B\,
F(\alpha,\beta,\gamma,z),
\end{eqnarray}
where $A$ and $B$ are arbitrary integration constants  and 
$F(\alpha,\beta,\gamma,z)\equiv {}_2F_1(\alpha,\beta;\gamma;z)$ is an ordinary 
hypergeometric function \cite{Abromowitz}. By considering (\ref{RFtrans}), the general solution of the
radial Klein-Gordon equation near the horizon can be written as 

\begin{eqnarray}
R&=&A\, 
z^{-i\tilde \omega}(1-z)^{\frac{1+\sqrt{1+4\tilde \zeta}}{2}} \, F(\alpha-\gamma+1,\beta-\gamma+1,2-\gamma,z)\nonumber\\
&+&B\, 
z^{i\tilde \omega}(1-z)^{\frac{1+\sqrt{1+4\tilde \zeta}}{2}}  \, F(\alpha,\beta,\gamma,z).
\end{eqnarray}
The first and second terms represent an ingoing and outgoing wave respectively 
at the horizon $z=0$. Because we are working at the classical level, there cannot 
be outgoing flux at the horizon. Therefore the second term of the solution  
should vanish, which requires $B=0$ and the radial part of the solution near horizon becomes
\begin{eqnarray}\label{horizonnearregion}
R&=&A\, 
z^{-i\tilde \omega}(1-z)^{\frac{1+\sqrt{1+4\tilde \zeta}}{2}}F\left(\frac{1+\sqrt{1+4\tilde \zeta}}{2},\frac{1+\sqrt{1+4\tilde \zeta}}{2}-2i\,\tilde \omega,1-2i\,\tilde \omega,z\right).
\end{eqnarray}

Since we will match the near and the far region solutions at the intermediate 
region, we need the large $r$, i. e., $z\rightarrow1$ limit of this solution. We can  use the 
 $z\rightarrow 1-z$ transformation law \cite{Abromowitz} for the hypergeometric functions
\begin{eqnarray}\label{transformation}
F(a,b,c,z)&=&\frac{\Gamma(c)\Gamma(c-a-b)}{\Gamma(c-a)\Gamma(c-b)}F(a,b,a+b-c+1,
1-z)\\
&+&(1-z)^{c-a-b}\frac{\Gamma(c)\Gamma(a+b-c)}{\Gamma(a)\Gamma(b)}F(c-a,c-b,
c-a-b+1,1-z).\nonumber
\end{eqnarray}

Note that this formula does not work for $c-a-b=\pm k$, where $k$ is an integer and for this case a transformation involving logarithmic terms is required \cite{Abromowitz}, which will make the matching with the far solution at intermediate regions impossible. For our case, however, since $a=(1+\sqrt{1+4\tilde\zeta})/2,\ b= (1+\sqrt{1+4\tilde\zeta})/2-2i \tilde \omega$ and $c=1-2i\tilde \omega$, we have $c-a-b=-\sqrt{1+4\tilde\zeta}$.  Since for any value of $l$, $\tilde{\zeta}$ is not an integer, contrary to the previous claim \cite{WangHerdeiro}, unlike static black holes,  the above transformation holds for any value of $l$ for rotating black holes. Thus, we do not end up with any logarithmic terms  in the above transformation.  We can also see this if we replace  $\zeta=l(l+n+1)+\epsilon$ in the above expression; we find that
\begin{equation}
 c-a-b=-\sqrt{1+4\tilde\zeta}\cong -\left[1+\frac{2(l+\bar\epsilon)}{n+1}\right]+O(\bar\epsilon^2), 
\end{equation}
where
\begin{equation}
\bar\epsilon=\frac{\epsilon}{2l+n+1}.
\end{equation}
 The resulting expression can be an integer if and only if $\bar\epsilon$ vanishes and $l/(1+n)$ is half integer. Here $\epsilon$ is the term denoting small corrections to the eigenvalues of angular equation due to the  rotation of the black hole and mass of the scalar field given in (\ref{lambda}) and (\ref{lambda1}). It is not an arbitrarily small parameter, its value  can be calculated in principle using (\ref{lambda}). In summary, unlike static black holes, we can use the transformation law (\ref{transformation}) for any value of $l$, keeping in mind that there are always small correction terms to be added  to the integer   values $l$, making the relevant expressions close 
to an integer up to a small correction term.

Thus using the above transformation, by using the property that $F(a,b,c,0)=1$, and considering that the Compton wavelength of the scalar field should be much larger than the black hole horizon implying $\tilde m r_h \ll 1$,  the large $r$ limit of the near region 
solution can be expressed as follows:
\begin{equation}\label{nearregion}
R(r)=A_1\, r^l+A_2\, r^{-l-n-1},
\end{equation}
 with
\begin{eqnarray}
&&A_1= \frac{A\, \Gamma(1-2i\tilde \omega)\, \Gamma\left( \frac{2l}{1+n}+1\right)}{\Gamma\left(\frac{l}{1+n}
+1\right)\Gamma\left(\frac{l}{1+n}+1-2i\tilde\omega\right)}
\left(r^{1+n}_+-r^{1+n}_-\right)^{-\frac{l}{1+n}},\\
&& A_2= \frac{A\, \Gamma(1-2i\tilde \omega)\, \Gamma\left(-1-\frac{2l+2\bar\epsilon}{1+n}\right)}{\Gamma\left(-\frac{l+\bar\epsilon}{1+n}
\right)\Gamma\left(-\frac{l}{1+n}-2i\tilde\omega\right)}
\left(r^{1+n}_+-r^{1+n}_-\right)^{1+\frac{l}{1+n}},
\end{eqnarray}
where we have also used the fact that 
$1-z\rightarrow\frac{r^{1+n}_+-r^{1+n}_-}{r^{1+n}}$ for $r\rightarrow\infty$ to derive the above expressions. For clarity we discarded  $\bar\epsilon$ terms in the above expressions except for the terms which might have poles if $\epsilon$ vanishing.  Actually, as we will see, the erroneous conclusion in \cite{AD} is not caused by the poles in the above equations but not taking these small corrections into the last step of the calculations.

\subsection{Far region solution}
For the far region $r-r_h\gg r_h$, we can ignore the effects of the black hole by setting $a=M=0$, then the radial equation reduces to \cite{WangHerdeiro}:
\begin{equation}
\left(1+\frac{r^2}{\ell^2}\right)\frac{d^2R}{dr^2}+ \left[(4+n)\frac{r}{\ell^2}+\frac{n+2}{r} \right]\frac{dR}{dr}+\left(\frac{\omega}{1+\frac{r^2}{\ell^2}}-\frac{\lambda}{r^2}  -\tilde{m}^2\right)R=0. 
\end{equation} 

By applying the transformations $y=1+r^2/\ell^2$ and $R(y)=y^{ \omega \ell/2} (1-y)^{l/2}\mathcal{F}(y)$ this equation reduces to a hypergeometric differential equation  of the form (\ref{hypergeometric}) 
 with
 \begin{eqnarray}
&&\alpha=\frac{n+3}{4}+\frac{l+\omega \ell}{2}+\frac{1}{2}\sqrt{\tilde{m}^2\ell^2+\left(\frac{n+3}{2}\right)^2} ,\\ 
&&\beta=\frac{n+3}{4}+\frac{l+\omega \ell}{2}-\frac{1}{2}\sqrt{\tilde{m}^2\ell^2+\left(\frac{n+3}{2}\right)^2}, \\ 
&& \gamma=1+\omega \ell.
\end{eqnarray} 

The regular solution of this equation,   obeying the boundary condition at $r=\infty$  is given by
\begin{equation}
R(y)=C (1-y)^{l/2}y^{\omega\ell/2-\alpha} F[\alpha,\alpha-\gamma+1,\alpha-\beta+1,y^{-1}],
\end{equation}
where $C$ is an integration constant. The small $r$ limit of this solution can be obtained by $\frac{1}{y} \rightarrow  1-y$  transformation of hypergeometric functions \cite{Abromowitz}, which yields 
\begin{eqnarray}\label{adsextsoln}
R(r)= C_1 r^l +C_2 r^{-l-n-1},
\end{eqnarray} 
 with
 \begin{eqnarray}
&& C_1=\frac{ C (-1)^{l/2}\, \Gamma\left[1+\sqrt{\tilde m^2 \ell^2 +\left(\frac{n+3}{2}\right)^2 } \right]\, \Gamma\left[-l-\frac{n+1}{2}\right] }{\ell^l\, \Gamma\left[ \frac{1-n}{4}-\frac{l+\omega \ell}{2}+\frac{1}{2} \sqrt{\tilde m^2 \ell^2 +\left(\frac{n+3}{2}\right)^2 } \right]\, \Gamma\left[ \frac{1-n}{4}-\frac{l-\omega \ell}{2}+\frac{1}{2} \sqrt{\tilde m^2 \ell^2 +\left(\frac{n+3}{2}\right)^2 } \right] },  \\
&& C_2= \frac{ C (-1)^{l/2}\, \Gamma\left[1+\sqrt{\tilde m^2 \ell^2 +\left(\frac{n+3}{2}\right)^2 } \right]\, \Gamma\left[l+\frac{n+1}{2}\right] }{\ell^{-(l+n+1)}\, \Gamma\left[ \frac{n+3}{4}+\frac{l+\omega \ell}{2}+\frac{1}{2} \sqrt{\tilde m^2 \ell^2 +\left(\frac{n+3}{2}\right)^2 } \right]\, \Gamma\left[ \frac{n+3}{4}+\frac{l-\omega \ell}{2}+\frac{1}{2} \sqrt{\tilde m^2 \ell^2 +\left(\frac{n+3}{2}\right)^2 } \right]}.\quad \quad
\end{eqnarray} 

Note that this solution (\ref{adsextsoln}) is for pure AdS spacetime for $r \in [0,\infty)$. The regularity of this solution at the origin requires the term $r^{-l-n-1}$ to vanish, which can be possible only if one of the $\Gamma$ functions  at the denominators of $C_2$ is equal to $\infty$. This yields a discrete (positive) spectrum \cite{WangHerdeiro}
\begin{equation}\label{frequency}
\omega_N \ell=2N+\frac{n+3}{2}+l+\sqrt{\tilde m^2 \ell^2 +\left(\frac{n+3}{2}\right)^2},
\end{equation}
 where $N=0,1,2,\ldots$ is the radial overtone number.  In the presence of the black hole horizon, it is natural to expect that the frequency of the waves has small (complex) modifications, compatible with  the limits we have considered in the  present analysis, as follows:
\begin{equation}
\omega=\omega_N+i \delta.
\end{equation}
Replacing this into $C_2$ and considering the fact that $
\Gamma(-N-i\delta\ell/2)^{-1}=(-1)^{N+1}\, i \delta\ell N!/2$, we have
\begin{eqnarray}
 C_2= \frac{ C (-1)^{l/2}\, (-1)^{N+1}i \delta N! \ \Gamma\left[1+\sqrt{\tilde m^2 \ell^2 +\left(\frac{n+3}{2}\right)^2 } \right]\, \Gamma\left[l+\frac{n+1}{2}\right] }{2\,\ell^{-(l+n+2)}\, \Gamma\left[ \frac{n+3}{2}+l+N+ \sqrt{\tilde m^2 \ell^2 +\left(\frac{n+3}{2}\right)^2 } \right]\,  }. 
\end{eqnarray} 
 
\subsection{Instability}
The near and far region solutions (\ref{nearregion}) and (\ref{adsextsoln}) can be matched at the intermediate region $r_h\ll r-r_h\ll 1/\omega$, since the asymptotic forms of these solutions have the same  powers of $r$. Considering the condition $C_1/A_1=C_2/A_2=1$,  the matching yields  the following expression for the correction term of the frequency
\begin{eqnarray}
&&\delta=2i \frac{(r_+^{1+n}-r_-^{1+n})^{1+\frac{2l}{1+n}}}{\ell^{2l+n+2}}\frac{(-1)^{N}}{N!}   
\frac{\Gamma\left(-1-\frac{2l+2\bar\epsilon}{1+n} \right)\Gamma\left(1+\frac{l}{1+n} \right)}{\Gamma\left(-\frac{l+\bar\epsilon}{1+n} \right)\Gamma\left(1+\frac{2l}{1+n} \right)}  \nonumber  \\
&&\quad \times \frac{\Gamma\left(-l-\frac{n+1}{2} \right) \Gamma\left(\frac{n+3}{2}+l+N+\sqrt{\tilde m^2\ell^2+\left(\frac{n+3}{2}\right)^2} \right)}{\Gamma\left(-\frac{n+1}{2}-l-N \right)\Gamma\left(l+\frac{n+1}{2} \right) \Gamma\left(1+N+\sqrt{\tilde m^2\ell^2+\left(\frac{n+3}{2}\right)^2} \right)}             
                 \frac{\Gamma \left(1+\frac{l}{1+n}-2i\tilde\omega\right)}{\Gamma\left(-\frac{l}{1+n}-2i\tilde\omega\right)}. 
\end{eqnarray} 
This main result is, apart from notational differences and the fact that the superradiance factors are different, in agreement with  the corresponding expression for the  charged static AdS black holes in arbitrary dimensions \cite{WangHerdeiro}. Although one can make a numerical analysis to investigate the presence of instability using the above expression, we can further expand the Gamma functions to prove it analytically. It turns out that the last multiplicative factor in the above expression is crucial to determine the real and imaginary parts of the correction to the frequency. Actually, we can use the expansion of the expression first presented by us in \cite{Turkuler} as
\begin{eqnarray}\label{ourtrn}
\frac{\Gamma \left(1+\frac{l}{1+n}-2i\tilde\omega\right)}{\Gamma\left(-\frac{l}{1+n}-2i\tilde\omega\right)}=&& -\frac{1}{\pi}\left|\Gamma\left(\frac{l}{1+n}+1-2i\tilde\omega \right)\right|^2 \nonumber \\
&& \times \quad \left\{\sin\left[\frac{(l+\bar{\epsilon})\pi}{1+n}\right]\cosh\left(2\pi\tilde\omega \right)+i \cos\left[\frac{(l+\bar\epsilon)\pi}{1+n}\right]\sinh\left(2\pi\tilde\omega \right) \right\}.
\end{eqnarray}
Note that the imaginary part of above equation vanishes if $\bar\epsilon=0$ and $l/(1+n)$ is half integer. This is the reason leading to the erroneous conclusion in \cite{AD} that in five dimensions, i.e. $n=1$, modes corresponding to odd values of $l$ do not trigger the instability. However, since $\bar\epsilon$ is not vanishing, the imaginary part of the above equation is also not vanishing, which corrects this erroneous conclusion. 

We can evaluate the other gamma functions  as given in \cite{WangHerdeiro}:
\begin{eqnarray}
&&\frac{\Gamma\left(-l-\frac{n+1}{2} \right)}{\Gamma\left(-l-\frac{n+1}{2}-N \right)}=(-1)^{N}\prod_{j=1}^N\left(l+\frac{n+1}{2}+j \right),\\
&& \frac{\Gamma\left(-1-\frac{2l+2\bar\epsilon}{1+n} \right)}{\Gamma\left(-\frac{l+\bar\epsilon}{1+n} \right)}= -\frac{1}{2 \cos \left[\frac{(l+\bar{\epsilon}) \pi }{1+n} \right] } \frac{\Gamma\left(1+\frac{l}{1+n} \right) }{\Gamma\left(\frac{2l}{1+n}+2 \right)}.\label{Herdeirotrn}
\end{eqnarray}
Using these formulas,  we can express the correction term as a sum of its real and imaginary parts as follows:
\begin{eqnarray}
\delta=-\sigma\left\{  \sinh\left(2\pi\tilde\omega \right)-i \tan\left[\frac{(l+\bar\epsilon)\pi}{1+n}\right]\cosh\left(2\pi\tilde\omega \right) \right\},
\end{eqnarray}
where the positive multiplicative factor $\sigma$ has the expression
\begin{eqnarray}
\sigma&=& \left|\Gamma\left(\frac{l}{1+n}+1-2i\tilde\omega \right)\right|^2 \frac{(r_+^{1+n}-r_-^{1+n})^{1+\frac{2l}{1+n}}}{\pi\, \ell^{2l+n+2} N!}\frac{\Gamma^2\left(1+\frac{l}{1+n} \right)}{ \Gamma\left(1+\frac{2l}{1+n} \right)\Gamma\left(2+\frac{2l}{1+n} \right)} \nonumber \\ 
&&\times \frac{\Gamma\left(\frac{n+3}{2}+l+N+\sqrt{\tilde m^2\ell^2+\left(\frac{n+3}{2}\right)^2} \right)  }{\Gamma\left(l+\frac{n+1}{2} \right) \Gamma\left(1+N+\sqrt{\tilde m^2\ell^2+\left(\frac{n+3}{2}\right)^2} \right)}\prod_{j=1}^N\left(l+\frac{n+1}{2}+j \right).
\end{eqnarray}
Thus the total frequency of the scalar field becomes
\begin{eqnarray}
\omega=\omega_N+i\delta=\omega_N-\sigma\tan \left[\frac{(l+\bar\epsilon)\pi}{1+n}\right]\cosh\left(2\pi\tilde\omega \right)-i \sigma  \sinh\left(2\pi\tilde\omega \right),
\end{eqnarray}
where $\omega_N$ is given in (\ref{frequency}). The real and imaginary parts of the frequency hence become
\begin{eqnarray}\label{RepartAds}
&&\mathfrak{Re}[\omega]=\omega_N-\sigma \tan \left[\frac{(l+\bar\epsilon)\pi}{1+n}\right]\cosh\left(2\pi\tilde\omega \right),\\
&&\mathfrak{Im}[\omega]=-\sigma \sinh{(2\pi\tilde\omega)}.\label{ImpartAdS}
\end{eqnarray}

The results presented above confirm that, irrespective of the value of $l$, the rotating AdS black holes always suffer from the superradiance instability whenever the frequency of the scalar field satisfies the superradiance condition $\tilde \omega<0$ which means $\omega-m \Omega_h<0$,
since we have
\begin{equation}
\Phi\sim e^{-i\omega t}=e^{-i \mathfrak{Re}[\omega] t}\, e^{\mathfrak{Im}[\omega]t}.
\end{equation}
 Thus, the imaginary part of the frequency is always positive under this condition, resulting in an exponential growth of the field. The time scale of this exponential growth is inversely proportional to the imaginary part of the frequency, i. e.,  $\tau\sim 1/\mathfrak{Im}[\omega]$. This fact allows us to obtain a relation between the AdS radius of the spacetime and the instability time scale as
\begin{equation} \label{timescaleAdS}
\tau\sim \ell^{2(l+1)+n}.
\end{equation}
Hence $\mathfrak{Im}[\omega]$ decreases and $\tau$ increases  with increasing $n$ for fixed AdS radius. This implies that 
as $D\rightarrow \infty$, $\tau\rightarrow \infty$ and the instability becomes ineffective in the large $D$ limit.  

The onset of the instability is determined by $\mathfrak{Re}[\omega]=m\Omega_h$. Since the real part of the frequency is inversely proportional to the AdS radius, there is a critical value of  $\ell_0$ such that a superradiant wave starts becoming nonsuperradiant. This particular value is $\ell_0=m\Omega_h/\omega_N$.

If we compare our results with the black hole bomb mechanism of Myers-Perry black holes \cite{CardosoBomb,Lee}, which is rederived in the Appendix for a comparison, we see that there is a perfect agreement with the corresponding expressions of real and imaginary parts of the frequency correction terms \{[(\ref{RepartAds}) and (\ref{ImpartAdS})] and [(\ref{realpart}) and (\ref{Impart})]\} and the time scale of the instability [(\ref{timescaleAdS}) and  (\ref{timescalebomb})].

Note that, although the above results will not change, some of the  gamma functions in  the above expressions can be further expanded  when the term  $l/(1+n)$ is integer or half integer. Since we have taken care of the possible poles in the gamma functions, we do not have to consider small correction terms to the $l$ below.

\subsubsection{The case $\frac{l}{1+n}=k$ where  $k$ is an integer} 
For this case, it is possible to further expand the general equation since  some of the gamma functions can be easily expanded and also we have the following relation:
\begin{eqnarray}
\left|\Gamma\left(\frac{l}{1+n}+1-2i\tilde\omega \right)\right|^2=\frac{2\pi \tilde\omega}{\sinh{(2\pi \tilde\omega)}}\prod_{p=1}^{k}\left(p^2+4 \tilde\omega^2\right).
\end{eqnarray}
The resulting expression for the frequency is
\begin{eqnarray}
\omega=\omega_N-\frac{\pi \bar\epsilon \sigma'}{1+n}\tilde\omega\,\coth{(2\pi\tilde\omega)}-i\sigma'\tilde\omega,
\end{eqnarray}
with
\begin{eqnarray}
\sigma'&=& 2 \frac{(r_+^{1+n}-r_-^{1+n})^{1+2k}}{\ell^{2l+n+2} N!}\frac{(k!)^2}{(2k)!(2k+1)!} \frac{\Gamma\left[\frac{n+3}{2}+l+N+\sqrt{\tilde m^2\ell^2+\left(\frac{n+3}{2}\right)^2} \right]  }{\Gamma\left(l+\frac{n+1}{2} \right) \Gamma\left[1+N+\sqrt{\tilde m^2\ell^2+\left(\frac{n+3}{2}\right)^2} \right]} \nonumber \\ 
&&\times\prod_{p=1}^{k}\left(p^2+4\tilde\omega^2 \right)\times\prod_{j=1}^N\left(l+\frac{n+1}{2}+j \right).
\end{eqnarray}
It is clear that the imaginary part of the frequency becomes positive under superradiance, implying the superradiance instability for small rotating  AdS black holes. These results  and the correction terms  are in accordance with the previous works  \cite{CardosoDias,AD1,Aliev5d,WangHerdeiro}. The real correction term is very small compared to $\omega_N$ and thus can be ignored.

\subsubsection{The case $\frac{l}{1+n}=k+\frac{1}{2}$ where  $k$ is an integer} 
For this case we have
\begin{eqnarray}
\left|\Gamma\left(\frac{l}{1+n}+1-2i\tilde\omega \right)\right|^2=\frac{\pi}{\cosh(2\pi\tilde\omega)}\prod_{p=1}^{k+1}\left[\left(p-\frac{1}{2}\right)^2+4\tilde\omega^2 \right],
\end{eqnarray}
and also
\begin{equation}
\Gamma{\left(1+\frac{l}{1+n}\right)}=2^{-k-1}\sqrt{\pi}\, (2k+1)!!\,.
\end{equation}
Using these expressions and expanding the trivial gamma functions, we have
\begin{equation}\label{freq12}
\omega=\omega_N+\frac{(1+n)\sigma'}{\bar\epsilon \pi}-i\sigma' \tanh{(2\pi \tilde\omega)}
\end{equation}
with
\begin{eqnarray}
\sigma'&=& \frac{\pi}{2^{2(1+k)}} \frac{(r_+^{1+n}-r_-^{1+n})^{2+2k}}{\ell^{2l+n+2} N!}\frac{[(2k+1)!!]^2}{(2k+1)!(2k+2)!} \frac{\Gamma\left[\frac{n+3}{2}+l+N+\sqrt{\tilde m^2\ell^2+\left(\frac{n+3}{2}\right)^2} \right]  }{\Gamma\left(l+\frac{n+1}{2} \right) \Gamma\left[1+N+\sqrt{\tilde m^2\ell^2+\left(\frac{n+3}{2}\right)^2} \right]} \nonumber \\ 
&&\times\prod_{p=1}^{k+1}\left[\left(p-\frac{1}{2}\right)^2+4\tilde\omega^2 \right]\times\prod_{j=1}^N\left(l+\frac{n+1}{2}+j \right).
\end{eqnarray}
This result shows that when $\frac{l}{1+n}$ is an half integer, which can only be possible when spacetime dimensions are odd, there is a superradiance instability when the wave satisfies the superradiance condition, because the imaginary part of the frequency becomes positive. This  result analytically corrects the erroneous claim in \cite{AD} that  for rotating AdS black holes in five dimensions there is no instability for even values of $l$. Hence, the above expressions for $D=5\ (n=1)$  show that there indeed exists an imaginary part of the frequency which is erroneously claimed to be vanishing in \cite{AD}. This claim was actually corrected in  \cite{WangHerdeiro} for charged static AdS black holes using only a numerical method, since the analytical method fails for those values of $l$ for static black holes. Thus, our paper completes the picture by showing analytically that the instability  exists for rotating AdS black holes with the help of the fact that the analytical methods are valid for any  value of $l$  for rotating black holes. 

Since the real correction term, the second term in (\ref{freq12}), involves a divergent term  $(1+n)$ for large $n$, we need to compare its magnitude with $\omega_N$ given in (\ref{frequency}) and also its behavior for large values of $n$. To analyze the ratio of the real correction term to normal modes, we consider the case when $\tilde m=k=r_-=0$ and $N=1$ for simplicity,  then we have 
\begin{equation}
\frac{(1+n)\sigma'}{\bar{\epsilon}\pi\, \omega_N}=\left(\frac{r_+}{\ell}\right)^{2(n+1)} \frac{(1+n)^2(2+n)\Gamma[4+l+n]}{2 \, \epsilon \, \,(11+3n)\, \Gamma[1+n]\,\Gamma\left[\frac{7+n}{2}\right]}.
\end{equation}
Note that since the small parameter  $\epsilon$ is not arbitrary  and  $(r_+/\ell)^{2(n+1)}/\epsilon$ is finite, this ratio is finite and in the limits we consider where $r_+\ll \ell$, the term $(r_+/\ell)^{2(n+1)}$ decays exponentially for large $n$, making the combination small. Hence the correction term is several orders of magnitude smaller than normal modes $\omega_N$, for example in five dimensions ($n=1$), for reasonable values $\ell=1$, $r_+=0.01$ and $a=0.001$, we have $\omega_N=7$ and the correction term  is $0.038$, which is $0.5\%$ of $\omega_N$, where we have chosen $\epsilon=(a \omega_N)^2$. Our analysis also shows that  when $k$ or $n$ increases this term becomes much smaller compared to $\omega_N$, keeping in mind that $\omega_N \ell=2N+n+3+l$  for massless case (\ref{frequency}). Also for large values of $n$, the above term gets much smaller and for the  $n\rightarrow \infty$ limit it converges to zero.

\section{Conclusion}

In this paper we have  investigated the superradiance instability of a scalar field for  singly rotating small AdS black holes in arbitrary dimensions analytically. This instability is originated from the existence of the superradiance mechanism of black holes together with the reflective boundary conditions of the asymptotic infinity of the AdS spacetime. As a consequence of this behavior, the boundary of the AdS spacetime behaves like an infinite potential barrier, localizing the scalar field near the horizon leading to an instability. The end point of this instability is expected such that we have a rotating hairy black hole having less energy and angular  momentum, living in an AdS spacetime whose boundary rotating with the speed of light. Thus, the next direction on this research topic might be to explore the existence and the properties of  these scalar hairy AdS black holes.

We have  analytically shown that for rotating small AdS black holes perturbed by a scalar field satisfying the superradiance condition, there is always superradiant instability, irrespective of the value of spacetime dimensions $D\ge 4$ and the value of orbital number $l>0$. Our result generalizes the superradiance instability of a scalar field for rotating AdS black holes to arbitrary dimensions. Actually, in the  pioneering work \cite{CardosoDias} proving the instability of  four dimensional small rotating AdS black holes under scalar perturbations, it was claimed that such instability must be present for higher dimensions as well. However, an analytical analysis for  rotating small AdS black holes has not been presented, except for $D=5$. For charged static AdS black holes, the instability is proved for generic dimensions \cite{WangHerdeiro} together with the observation that, unlike for the rotating AdS black holes as we have shown in  our paper here,  the analytical method fails for certain values of the orbital quantum number.  The time scale of the instability is  proportional to a power of  the radius of the AdS spacetime increasing  with spacetime dimensions, similar to the case of the black hole bomb mechanism. These results are in accordance with  the instability of rotating black holes surrounded by a  hypothetical reflective mirror in the black hole bomb mechanism, in which we have reviewed at the Appendix. We have also recovered in the Appendix the previous results that the bosonic and fermionic thermal factors for a scalar field absorption  naturally arise for rotating black holes for $D=5$.

\section*{ACKNOWLEDGEMENTS}
The authors would like to thank Ahmet Baykal for reading the manuscript and useful discussions. O.D. and T. D. are supported by Marmara University Scientific Research Projects Committee (Project No: FEN-C-YLP-130313-0081).

%
%

\section*{APPENDIX: BLACK HOLE BOMB MECHANISM FOR MYERS-PERRY BLACK HOLES}

\subsection{Myers-Perry Black Hole Spacetime}
In order to compare the results we have found  for rotating AdS black holes in $D$ dimensions with the rotating black holes in black hole  bomb formalism, here we review the black hole bomb mechanism for $D$ dimensional rotating Myers-Perry black holes, which were studied before in four \cite{CardosoBomb} and generic dimensions \cite{Lee} and shown to  suffer from superradiance instability. Since the existence of the notational differences and the real and imaginary parts of the frequency correction terms are not calculated explicitly in \cite{Lee}, here we have preferred to  derive  all the relevant expressions  in detail. Hence here there are no new results except the explicit expressions of frequency correction terms, and the relations of decay rates with bosonic and fermionic thermal factors. 
This part is partially  based on \cite{Turkuler} with some corrections.     
We consider the $D=4+n$ dimensional Myers-Perry black hole with mass $M$ and a   single rotation parameter $a$ given by the metric
\begin{eqnarray}
ds^2=&-&\frac{\Delta-a^2\sin^2\theta}{\Sigma}dt^2
-\frac{2a(r^2+a^2-\Delta)}{\Sigma}\sin^2\theta dt d\phi
+\frac{(r^2+a^2)^2-\Delta\, a^2 \sin^2\theta}{\Sigma}\sin^2\theta\, 
d\phi^2\nonumber\\
&+&\frac{\Sigma}{\Delta}dr^2
+\Sigma d\theta^2
+r^2 \cos^2 \theta d\Omega^{2}_{n},
\end{eqnarray}
where the metric functions are given by
\begin{eqnarray}
\Sigma&=&r^2+a^2\cos^2\theta,\nonumber\\
\Delta&=&r^2+a^2-2Mr^{1-n},\label{Delta}
\end{eqnarray}
and here again $d\Omega^{2}_{n}$ is the standard  metric of the $n$-sphere. This black hole  has physical mass $\mu$ and angular momentum $J$ 
which can be found by setting $\Xi=1$  in (\ref{MJ}) for the same parameters for AdS black holes.
The event horizon of the black hole is located on the largest real root of $\Delta=0$ 
given  by
\begin{equation}
r_h^2+a^2-2M r_{h}^{1-n}=0.
\end{equation}

 In four and five dimensions the existence of the horizon sets an upper bound (extremal limit) on the rotation parameter $a$ ($a\le M$ for $D=4$ and and $a\le \sqrt{2M}$ for $D=5$). In these limits in four dimensions there are two horizons $r_{\pm}$ where they are called  the outer and the inner horizons. In dimensions greater than 4, however, there is always one horizon ($r=r_h$) and they can have an arbitrarily high rotation parameter $a$   if $D>5$. They are called  ultraspinning black holes and they have a regular horizon and arbitrarily large angular momentum per unit mass. These ultraspinning black holes are known to suffer from several instabilities \cite{UltraSpinning1,UltraSpinning2}. In this paper we consider the slow rotation regime, in which the ultraspinning instability is not present.

 The temperature of this black hole is given by \cite{EmparanReall}
\begin{equation}
T_h=\frac{1}{4\pi}\left( \frac{2\, r_h}{r_h^2+a^2}+\frac{D-5}{r_h}\right)\approx  \frac{1}{4\pi}\left(\frac{n+1}{r_h}\right),
\end{equation}
where the last expression is valid in the slow rotation limit $a \ll 1.$

\subsection{Klein-Gordon equation}

 Here we calculate the massless ($\tilde m=0$) Klein-Gordon equation for a scalar field $\Phi$  given by
\begin{equation}
\nabla^\mu \nabla_\mu \Phi=0.
\end{equation}
If we use the standard separation ansatz
\begin{equation}
\Phi=e^{im\phi-i \omega t}Y(\Omega)\Theta(\theta)R(r),
\end{equation}
we see that the Klein-Gordon equation is separated into angular and radial 
equations given by
\begin{eqnarray}\label{angular1}
\frac{1}{\cos^n\theta \sin\theta}\frac{d}{d\theta}\left(\cos^n\theta 
\sin\theta \frac{d\Theta}{d\theta}\right)+
\bigg[a^2\omega^2\cos^2\theta-\frac{m^2}{\sin^2 
\theta}-\frac{j(j+n-1)}{\cos^2 \theta}+\bar{\lambda}_{jlm}\bigg]\Theta=0,\quad\ \
\end{eqnarray}
\begin{eqnarray}\label{radial1}
\frac{1}{r^{n}}\frac{d}{dr}\left(\Delta r^n \frac{dR(r)}{dr}\right)
+\bigg[\frac{(a^2+r^2)^2}{\Delta}\left(\omega-\frac{ma}{a^2+r^2}\right)^2-\frac{
j(j+n-1)a^2}{r^2}-\tilde{\lambda}_{jlm}\bigg]R(r)=0,\quad
\end{eqnarray}
where the constants $\bar{\lambda}_{jlm}$ are given by 
 \begin{eqnarray}
 \tilde{\lambda}_{jlm}=\bar{\lambda}_{jlm}+a^2 \omega^2-2 a m \omega.
  \end{eqnarray}
As in the AdS case,  the terms $j(j+n-1)$ are eigenvalues of the spheroidal equation for $Y$ of the $n$-sphere \cite{Muller} where $j$ is an integer. For this case,  the angular equation (\ref{angular1}) takes the form   discussed in \cite{berti} and in the slow rotation limit we can expand the eigenvalues in a Taylor series of the following form  
\begin{equation}
\bar{\lambda}_{jlm}=l(l+n+1)+\sum_{p=1}^{\infty} f_p (a\omega)^p,
\end{equation}
 and its first several values are explicitly calculated in \cite{berti}. 

We can also express the radial equation in a Schr\" odinger--like form,
\begin{equation}
 \frac{d^2\mathcal{R}}{dr^{*2}}+V(r)\mathcal{R}=0,
 \end{equation}
 where transformation equations and the potential term  are given in (\ref{tortR}), (\ref{tortr}), and (\ref{tortpot})  by setting $\Xi=1$, $\tilde m=0$ and replacing $\Delta_r$ with $\Delta$ given in (\ref{Delta}).
Near the horizon, the behavior of the scalar field is the same with the AdS case (\ref{potnearhor}), i. e., $V(r\rightarrow r_h)\sim (\omega-m \Omega_h)^2$ with asymptotic behavior of the field $R(r\rightarrow r_h)\sim e^{-i\omega t-i(\omega-m \Omega_h)r^*}$, where the angular  velocity of the horizon for this case is given by 
\begin{equation}
 \Omega_h=\frac{a}{r^2_h+a^2}.
\end{equation}

The asymptotic form of the radial function at radial infinity is different from the AdS case, since, for $r\rightarrow \infty$ the potential becomes $V(r\rightarrow 
\infty)=\omega^2$, which yields the solution to be of the form $R\sim e^{\pm 
i\omega r^*}$. It is easy to verify that when the well--known condition,
\begin{equation}
\omega-m\Omega_h <0,
\end{equation}
is satisfied  the waves reflected back from the black hole with increased
energy and the phenomena called  superradiance occurs. 
The question is 
whether it is possible to localize waves satisfying this condition near the horizon to lead to an 
instability on the black hole. One of the methods for localizing these waves is  considering an artificial reflecting mirror surrounding the black hole, i. e., black hole bomb mechanism,  in   which  the waves are continuously scattered between the black hole horizon and the mirror.  Since we want to compare the results we have found in the previous section for rotating AdS black holes and instability of rotating black holes surrounded by a reflecting mirror found  before  in the four dimensional case \cite{CardosoBomb}, and for generic dimensions in \cite{Lee}, here we investigate the stability of Myers-Perry black holes under  black hole bomb mechanism. To use the asymptotic matching technique, we will need the near and far region solutions of the Klein-Gordon equation.

\subsection{Near-region and far region solutions}
Since near the horizon the Myers-Perry and  rotating AdS black holes, in the slow rotation low frequency limit, have similar properties, the Klein-Gordon equation  near the horizon has the same form (\ref{radnearads}). Thus, following similar steps, the near horizon solution and its far region extension also have the same forms given in (\ref{horizonnearregion}) and (\ref{nearregion}). The difference will be at the far region solution in which we will now solve in detail.

In the far region, $r-r_+\gg M$, since the effects of the black hole can be 
neglected we have $a\sim 0, M\sim0, \Delta\sim r^2$. Therefore, the radial 
equation is reduced to
\begin{eqnarray}
\frac{d^2 R(r)}{dr^2}
+\frac{n+2}{r}\frac{dR(r)}{dr}
+\left[\omega^2-\frac{\tilde\lambda_{jlm}}{r^2}\right]R(r)=0.
\end{eqnarray}
The general solution of this differential equation can be 
expressed in the form of a linear combination of Bessel functions of the first 
and the second kind $J$ and  $Y$ as
\begin{eqnarray}\label{Bessel}
R(r)=\frac{1}{r^{\frac{1+n}{2}}}\left[\alpha\, J_{l+\frac{n+1}{2}}(\omega r)+\beta \,
Y_{l+\frac{n+1}{2}}(\omega r)\right],
\end{eqnarray}
or equivalently in terms of Hankel functions 
\begin{eqnarray}\label{Hankel}
R(r)=\frac{1}{r^{\frac{1+n}{2}}}\left[D_1 H^{(1)}_{l+\frac{1+n}{2}}(\omega 
r)+D_2 H^{(2)}_{l+\frac{1+n}{2}}(\omega r)\right],
\end{eqnarray}
by considering the identities relating Bessel functions to Hankel functions   \cite{Abromowitz}.
Here, the relations between the integration constants of both forms of the solutions  are given by
\begin{equation}
D_1=\frac{\alpha-i\beta}{2},\quad D_2=\frac{\alpha+i\beta}{2}.
\end{equation}

In order to match this far region solution with the near region solution at intermediate regions, we will need the small $r$ behavior of the far region solution, which can be easily found from (\ref{Bessel}) by using the asymptotic forms of the Bessel functions at small values \cite{Abromowitz}.
Then, the small $r$ limit of  the far region solution becomes
\begin{eqnarray}\label{farregion}
R(r)\sim-\left(\frac{\omega}{2}\right)^{-l-\frac{1+n}{2}}\frac{\beta}{\pi}
\Gamma\left(l+\frac{1+n}{2}\right)\, r^{-(1+n+l)}
+\left(\frac{\omega}{2}\right)^{l+\frac{1+n}{2}}\frac{\alpha}{
\Gamma\left(l+\frac{1+n}{2}+1\right)}\,r^l.
\end{eqnarray}

We will also need $r \rightarrow \infty$ behavior of the far region solution, which can be easily derived by considering (\ref{Hankel}) form of the solution with the asymptotic behavior of Hankel functions for large values of their arguments as given in \cite{Abromowitz}. Then the asymptotic form of the far region solution becomes  
\begin{eqnarray}\label{infinityfarregion}
R(r\rightarrow \infty)\sim\sqrt{\frac{2}{\pi \omega}}\frac{1}{r^{\frac{2+n}{2}}}
\left[D_1 e^{i(\omega r-\frac{l\pi}{2}-\frac{(1+n) \pi}{4}-\frac{\pi}{4})}
+D_2 e^{-i(\omega r-\frac{l\pi}{2}-\frac{(1+n) \pi}{4}-\frac{\pi}{4})}\right].
\end{eqnarray}

\subsection{Matching near and far region solutions}

The matching of the near region and far region solutions requires the powers of $r$ of the distinct parts to  solutions  to be the same for  both solutions. 
 When $M\ll r-r_+ \ll\frac{1}{\omega}$, the near region solution 
(\ref{nearregion}) as $r\rightarrow\infty$ and the far region solution 
(\ref{farregion}) as $r\rightarrow0$ overlap and this matching yields
\begin{eqnarray}\label{Aalpha}
\frac{A}{\alpha}=\left(\frac{\omega}{2}\right)^{l+\frac{1+n}{2}}(r^{1+n}_+-r^{
1+n}_-)^{\frac{l}{1+n}}
\frac{\Gamma\left(\frac{l}{1+n}+1\right)\Gamma\left(\frac{l}{1+n}
+1-2i \tilde\omega\right)}{\Gamma\left(l+\frac{1+n}{2}+1\right)\Gamma\left(\frac{2l}{
1+n}+1\right)\Gamma\left(1-2i\tilde\omega\right)},
\end{eqnarray}
and
\begin{eqnarray}\label{betaalfa1}
\frac{\beta}{\alpha}&=&-\pi\left(\frac{\omega}{2}\right)^{2l+n+1}(r^{1+n}_+-r^{
1+n}_-)^{1+\frac{2l}{1+n}}
\frac{1}{\Gamma\left(l+\frac{1+n}{2}+1\right)\Gamma\left(-\frac{l+\bar{\epsilon}}{1+n}\right)}
\nonumber\\
&\times&\frac{\Gamma\left(\frac{l}{1+n}+1\right)}{\Gamma\left(\frac{2l}{1+n}
+1\right)}
\frac{\Gamma\left(-\frac{2(l+\bar{\epsilon})}{1+n}-1\right)}{\Gamma\left(l+\frac{1+n}{2}\right)}
\frac{\Gamma\left(\frac{l}{1+n}+1-2i\tilde \omega\right)}{\Gamma\left(-\frac{l}{1+n}
-2i\tilde\omega\right)}.
\end{eqnarray}

\subsection{Scalar field absorption and decay rates}
Let us calculate here the absorption cross sections and decay rates of  a massless scalar field from a rotating black hole with single spin. Note that these were investigated before especially in the context of brane world picture in \cite{Creek}.

\subsubsection{The flux and the absorption cross section of the scalar field}

We can calculate the flux using the formula
\begin{eqnarray}
J=\frac{r^n\Delta }{2i}
\bigg[R^\ast  \frac{dR}{dr} - R \frac{dR^\ast}{dr} \bigg],
\end{eqnarray}
where * denotes complex conjugation. The incoming and outgoing fluxes  at radial infinity are calculated as
\begin{eqnarray}
J_{in}&=&-\frac{2}{\pi} |D_2|^2,\nonumber\\
J_{out}&=&\frac{2}{\pi} |D_1|^2.
\end{eqnarray}
The absorbed flux from the black hole horizon located at $z=0$ can be calculated from the near region solution of the form (\ref{horizonnearregion}) using the appropriate formula 
\begin{eqnarray}
J=\frac{(n+1)(x_+-x_-)\,z}{2i}
\left(R^\ast \frac{dR}{dz} - R  \frac{dR^\ast}{dz} \right),
\end{eqnarray}
which yields
\begin{eqnarray}
J_{abs}=-(n+1)(x_+-x_-)\,\tilde\omega\, |A|^2=-(\omega-m \Omega_h)r_+^{n+2}|A|^2,
\end{eqnarray}
where we have used  the definition of $\tilde{\omega}$ given in (\ref{omegatilde}) at the last step.
The absorption probability is given by
\begin{eqnarray}\label{absprob}
1-|S_l|^2&=&\frac{J_{abs}}{J_{in}} = (\omega-m \Omega_h)\, \frac{\pi\,r_+^{n+2}}{2}   \frac{|A|^2}{|D_2|^2}.
\end{eqnarray}
 
Now we are ready to calculate the absorption cross section for a massless scalar field from  a $D=4+n$ dimensional singly rotating black hole using the formula \cite{gubser}
\begin{eqnarray}
\sigma_l&=&\frac{2^{1+n}\pi^{\frac{1+n}{2}}}{\omega^{2+n}}\,\Gamma\left(\frac{1+n}{2}\right)\left(l+\frac{1+n}{2}\right)
\begin{pmatrix}
l+n\\
l
\end{pmatrix}
\left(1-|S_l|^2\right).
\end{eqnarray}

Using the expression (\ref{absprob}) above, we have 
\begin{eqnarray}
\sigma_l=(\omega-m \Omega_h)\,r_+^{n+2}\, \frac{2^n\pi^{\frac{3+n}{2}}}{\omega^{2+n}}\,\Gamma\left(\frac{1+n}{2}\right)\left(l+\frac{1+n}{2}\right)
\frac{(l+n)!}{n! l!}\frac{|A|^2}{|D_2|^2}.
\end{eqnarray}
Since for small frequency limit $\beta\ll \alpha$, we can approximate  $|A|^2/|D_2|^2\approx 4|A/ \alpha|^2$, and using (\ref{Aalpha}), the absorption cross section becomes
\begin{eqnarray}\label{nsigmaabs}
\sigma_l&=& \left(\omega-m\Omega_h \right)
\frac{\pi^{\frac{3+n}{2}}}{2^{2l-1}}\omega^{2l-1}
r_+^{2+n}(r^{1+n}_+-r^{1+n}_-)^{\frac{2l}{1+n}}
\left(l+\frac{1+n}{2}\right)
\frac{(l+n)!}{n!\, l!}\nonumber\\
&& \times\ \Gamma\left(\frac{1+n}{2}\right)
\bigg|\frac{\Gamma\left(\frac{l}{1+n}+1\right)\Gamma\left(\frac{l}{1+n}+1-2i\tilde\omega\right)}
{\Gamma\left(l+\frac{1+n}{2}+1\right)\Gamma\left(\frac{2l}{1+n}+1\right)\Gamma\left(1-2i\tilde\omega \right)}\bigg|^2.
\end{eqnarray}
Thus, we see that in the case of superradiance, $\omega-m\Omega_h<0$,  the absorption cross section becomes negative implying the amplification of the scalar field or equivalently the energy extraction  from the black hole.

\subsubsection{Decay rates}
Now let us calculate the decay rate of this scalar field using the formula
\begin{eqnarray}\label{ndecay}
\Gamma_l=\frac{\sigma_l}{e^{\beta_H (\omega-m \Omega_h )}-1},
\end{eqnarray}
where $\beta_H$ is the inverse Hawking temperature of the black hole and in the slow rotation limit it is given by (remember that $r_+=r_h$ and $r_-=0$ for $D > 4$)
\begin{eqnarray}\label{nbetah}
\beta_H=\frac{4\pi}{1+n} \frac{r_+^{2+n}}{r_+^{1+n}-r_-^{1+n}}.
\end{eqnarray}
The general equation of this expression can be obtained   by replacing  $\sigma_l$  in (\ref{ndecay}) with its corresponding  expression (\ref{nsigmaabs}). Now let us calculate its special values when the ratio $\frac{l}{1+n}$ is integer or half integer.

When the ratio $\frac{l}{1+n}=k$ is an integer, the decay rate becomes
\begin{eqnarray}
\Gamma_l=&& \frac{ \left(\omega-m\Omega_h \right)\omega^{2l-1}}{e^{\beta_H (\omega-m\Omega_h)}-1} \,
\frac{\pi^{\frac{3+n}{2}}}{2^{2l-1}}
\, r_+^{2+n}(r^{1+n}_+-r^{1+n}_-)^{\frac{2l}{1+n}}
\left[\frac{k!}{(2k)!}\right]^2\nonumber\\
&& \times \left(l+\frac{1+n}{2}\right)
\frac{(l+n)!}{n! l!}
\frac{\Gamma\left(\frac{1+n}{2}\right)}{\Gamma^2\left(l+\frac{1+n}{2}+1\right)}
\prod_{j=1}^{k}\left(j^2+4\tilde\omega^2 \right).
\end{eqnarray}
This corresponds to the decay rate of a bosonic field. For $n=0$ this expression reduces to the decay rate of the Kerr black hole. It is also compatible with the decay rates of five dimensional black holes if the orbital quantum number of the scalar field is even \cite{Maldacena,DiasEmparan,AD1}. These bosonic  terms arise in all dimensions whenever $l/(1+n)$ is an integer.

When the ratio  $\frac{l}{1+n}=k+1/2$  is half integer where $k$ is again an integer, which is only possible for $D>4$ and for odd $D$, we can also expand the gamma functions in the general expression of decay rate. Following the similar steps, we find that
\begin{eqnarray}
\Gamma_l=&&\frac{ \omega^{2l-1}}{e^{\beta_H (\omega-m\Omega_h)}+1}\, \frac{\pi^{\frac{5+n}{2}} }{2^{2(l+k+1)}}(r^{1+n}_+-r^{1+n}_-)^{\frac{2l}{1+n}+1}
(n+1)\left(l+\frac{1+n}{2}\right)
\frac{(l+n)!}{n! l!}
\left[\frac{(2k+1)!!}{(2k+1)!}\right]^2
\nonumber\\
&&\times\frac{\Gamma\left(\frac{1+n}{2}\right)}{\Gamma^2\left(l+\frac{1+n}{2}+1\right)}
\prod_{j=1}^{k+1}\left[\left(j-\frac{1}{2}\right)^2+4\tilde\omega^2 \right].
\end{eqnarray}

This decay rate is similar to the decay rate of a fermionic field. These expressions agree for the decay rates for five dimensional black holes \cite{Maldacena,DiasEmparan,AD1}. In five dimensions since we can have only integer or half integer values of $l/(1+n)$ the decay rates are either bosonic for even $l$ or fermionic for odd $l$.   
For dimensions greater than 5, however, $l/(1+n)$ is not an integer or half integer except for some special cases, the decay rates are not either purely fermionic or purely bosonic.  These bosonic and fermionic terms, which imply the existence of a conformal symmetry, are  also obtained in the discussion about high $D$ limit of general relativity in \cite{Emparan}. The possibility of whether the general decay rate expression (\ref{ndecay}) corresponds to  a combination of both factors requires further investigations.

\subsection{Black hole bomb and superradiance instability}

Now we surround the $D$ dimensional singly rotating black hole with an artificial mirror having reflective walls as done in \cite{CardosoBomb,Lee}. The radius of this wall must be large in order 
to use the approximations we have adopted. At the location of the mirror, $r=r_0$, the radial 
part of the wave function must vanish, i.e.
\begin{equation}
R(r_0)=0.
\end{equation} 

Considering this condition on the far solution (\ref{Bessel}), we have
\begin{equation}\label{alphabeta}
\frac{\beta}{\alpha}=-\frac{J_{l+(n+1)/2}(\omega r_0)}{Y_{l+(n+1)/2}(\omega 
r_0)},
\end{equation}
where the left--hand side of this equation is given in (\ref{betaalfa1}).

When there is no black hole, the scalar field will develop stable modes which may be labeled by $\omega=\omega_0$. The interaction of the scalar field with the black hole will affect this configuration and the  frequency of the scalar field  will be changed. We assume that, in the limits we consider, this change will be small. Thus, we consider that in the presence of the black hole the correction to the 
frequency of these modes will be of the form
\begin{equation}
\omega=\omega_0+i \delta, 
\end{equation}
where the absolute value of the correction term $\delta$ is assumed to be much 
smaller  than $\omega_0$.
Our aim is to determine the correction term $i\delta$.  
Under the small frequency limit, since left--hand side of the $(\ref{betaalfa1})$ is 
proportional to $\omega^{2l+n+1}$, we can take in the first approximation that 
the ratio $\beta/\alpha$ in (\ref{betaalfa1})  is vanishing. By considering 
(\ref{alphabeta}), for a given $\omega$ and $r_0$, we have
\begin{equation}\label{Jroots}
J_{l+\frac{n+1}{2}}(\omega_0 r_0)=0,
\end{equation}  
 which has solutions 
 \begin{equation}
\omega_0 r_0=j_{l+\frac{n+1}{2},N}, 
 \end{equation}
 where $j_{l+\frac{n+1}{2},N}$ are the roots of the equation (\ref{Jroots}). Their values can be found in \cite{Abromowitz} or can be obtained by using an analytical computer program. 
In this approximation, the frequency of the scalar field becomes
\begin{equation}
\omega=\omega_0+i \delta = \frac{j_{l+\frac{n+1}{2},N}+i\tilde\delta}{r_0}.
\end{equation} 
 
 Using the Taylor expansion in these approximations we have 
$J_{l+\frac{n+1}{2}}(\omega r_0)\simeq i \tilde \delta J_{l+\frac{n+1}{2}}'( 
j_{l+\frac{n+1}{2},N})$ in (\ref{alphabeta}) where $\delta=\tilde \delta/r_0$. 
Considering all these, from (\ref{alphabeta}) we find that
 \begin{eqnarray}\label{delta1}
 \tilde{\delta}=-i \pi\gamma \frac{Y_{l+\frac{n+1}{2}}(\omega_0 
r_0)}{J'_{l+\frac{n+1}{2}}(\omega_0 
r_0)}\frac{\Gamma\left(-1-\frac{2l+2\bar\epsilon}{1+n}\right)}{\Gamma\left(-\frac{l+\bar\epsilon}{1+n}\right)} 
\frac{\Gamma(1+\frac{l}{1+n}-2i \tilde  \omega)}{\Gamma(-\frac{l}{1+n}-2i\tilde\omega)}, 
\end{eqnarray}
where the factor
\begin{equation}
\gamma=\left(\frac{\omega_0}{2}\right)^{2l+n+1} 
\frac{\Gamma(1+\frac{l}{1+n})\, \left(r^{1+n}_+-r^{1+n}_-\right)^{1+\frac{2l}{1+n}}   }{\Gamma(1+\frac{2l}{1+n})\Gamma(1+l+\frac{1+n}{2}
)\Gamma(l+\frac{1+n}{2}) },
\end{equation}
is positive. Using the expansion of gamma functions  (\ref{ourtrn}) first given by us  in  \cite{Turkuler} and  the expansion(\ref{Herdeirotrn}) given in \cite{WangHerdeiro},  $\tilde \delta$ becomes
\begin{eqnarray}\label{delta2}
 \tilde{\delta}=-\mathcal{K} 
\left[\sinh{(2\pi\tilde\omega)}-i 
\tan \left[\frac{(l+\bar\epsilon)\pi }{1+n}\right]   \cosh{(2\pi\tilde\omega)} \right],
\end{eqnarray} 
  where 
 \begin{eqnarray}\label{K}
 \mathcal{K}= \frac{\gamma}{2}\frac{\Gamma\left(1+\frac{l}{1+n}\right)}{ \Gamma\left(2+\frac{2l}{1+n}\right)} \left|\frac{Y_{l+\frac{n+1}{2}}(\omega_0 
r_0)}{J'_{l+\frac{n+1}{2}}(\omega_0 
r_0)}\right|  
 \left|\Gamma(1+\frac{l}{1+n}-2i \tilde\omega)\right| 
^2.
\end{eqnarray}

 Thus the perturbation term $\tilde{\delta}$, hence the 
frequency of scalar field, has both real and imaginary parts. In the general case, i.e., if none of these terms  vanish, if the frequency of the scalar field satisfies the superradiance condition, and if the imaginary part is positive, both an instability and frequency shifts occur  for a black hole surrounded by a reflective 
mirror. This is in accordance with the previous works for generic dimensions \cite{Lee,Turkuler} and also for rotating charged black holes  in five dimensions \cite{AD1,Aliev5d}.
Since the frequency is given by $\omega=\omega_0+i \delta$, the real and imaginary parts of the frequency are
\begin{eqnarray}\label{realpart}
&&\mbox{Re}[\omega]=\omega_0-\mbox{Im}[\delta]=w_0-\frac{\mathcal{K}}{r_0}  \, \tan\left[\frac{(l+\bar\epsilon)\pi}{1+n}\right]   \cosh{(2\pi\tilde\omega)},
 \\
&&\mbox{Im}[\omega]=\mbox{Re}[\delta]=-\frac{\mathcal{K}}{r_0} 
\sinh{(2\pi\tilde\omega)}. \label{Impart}
\end{eqnarray}
The vanishing of the imaginary part (\ref{Impart}) implies that the scalar field does not show an instability and the perturbation is oscillating with time.  As it is obvious, this is not the case and instability is always present if the superradiance condition is satisfied.  Note that one can expand the $\Gamma$ functions in (\ref{delta2}) for some special cases of $l/(1+n)$ as in the AdS case, without altering the conclusions holding in the general case.

The time scale of the instability is $\tau=1/\mbox{Im}[\omega]=1/\mbox{Re}[\delta]$. Since one can deduce the relation $\delta\sim 1/{r_0^{2(l+1)+n}}$ from the above expressions  we see that the time scale increases with increasing mirror radius and also increasing spacetime dimensions, since we have  
\begin{equation}\label{timescalebomb}
\tau\sim r_0^{2(l+1)+n}.
\end{equation}
This implies that in the large $D$
limit, the instability becomes ineffective.

The real part of the  frequency of the waves is inversely proportional to the mirror radius $\omega\sim 1/r_0$, implying a minimum  mirror radius to instability to take place. The end point of the instability  is when the system reaches the critical frequency condition $\omega_c=m \Omega_h$, resulting in a black hole with smaller energy and angular momentum.

  \end{document}